\documentclass[prl,showpacs,twocolumn,floatfix]{revtex4}

\usepackage{amsmath,amssymb}
\usepackage{graphicx}
\usepackage[colorlinks]{hyperref}

\newcommand {\ch}  {{\rm ch}}

\newcommand {\E}  {{\varepsilon}}
\newcommand {\om}  {{\omega}}
\newcommand {\Om}  {{\Omega}}

\newcommand {\Ld} {{L_{\rm d}}}

\newcommand {\Lrad} {{L_{\rm r}}}

\newcommand {\Nd} {{N_{\rm d}}}

\newcommand {\dUmax} {U^{\prime}_{\max}}

\begin{document}

\title{Feasibility of an electron-based crystalline undulator}

\author{M.~Tabrizi}
\author{A.V.~Korol}
\altaffiliation[Also at: ]
{Department of Physics,
St Petersburg State Maritime Technical University,
St Petersburg, Russia}
\email[E-mail address: ]{a.korol@fias.uni-frankfurt.de}
\author{A.V.~Solov'yov}
\email[E-mail address: ]{solovyov@fias.uni-frankfurt.de}
\author{Walter Greiner}
\affiliation{
 Frankfurt Institute for Advanced Studies,
 Johann Wolfgang Goethe-Universit\"at,
Max-von-Laue-Str.~1, 60438 Frankfurt am Main, Germany}

\begin{abstract}
The feasibility to generate powerful monochromatic radiation of the 
undulator type in the gamma region of the spectrum
 by means of planar channeling of ultra-relativistic {\em electrons} in 
a periodically bent crystal is proven. 
It is shown that an electron-based crystalline undulator
operates in the regime of higher beam energies than a positron-based one does.
A numerical analysis is performed for a 50 GeV electron
channeling in Si along the (111) crystallographic planes.
\end{abstract}


\maketitle
 
In this Letter we demonstrate, for the first time, 
that it is possible to construct a powerful source of 
high energy photons by means of planar channeling of ultra-relativistic 
electrons through a periodically bent crystal.
For a positron channeling the feasibility of such a device was
demonstrated in Ref.~\cite{KSG1998_1999}.

A periodically bent crystal together with a bunch of ultra-relativistic 
charged particles which undergo planar channeling constitute a 
crystalline undulator. 
In such a system there appears, in addition to the well-known channeling 
radiation, the  undulator type radiation 
which is due to the periodic motion of channeling particles which follow 
the bending of the crystallographic planes \cite{KSG1998_1999}. 
The intensity and characteristic frequencies of this 
radiation can be varied by changing the beam energy 
and the parameters of the bending.
In the cited papers as well as in subsequent publications
(see the review \cite{KSG2004_review} and the references therein)
we proved a feasibility to create a short-wave crystalline undulator 
that will emit high-intensity, highly monochromatic radiation when pulses of 
ultra-relativistic 
{\em positrons} are passed through its channels. 
More recently it was demonstrated \cite{SPIE1} that 
the brilliance of radiation from a positron-based undulator 
in the energy range from hundreds of keV up to tens of MeV
is comparable to that of conventional light sources of the third 
generation operating for much lower photon energies.
Experimental study of this phenomenon is on the way within the framework 
of the PECU project 
\cite{PECU}.
\begin{figure}[ht]
\includegraphics*[width=8.5cm,height=3.5cm]{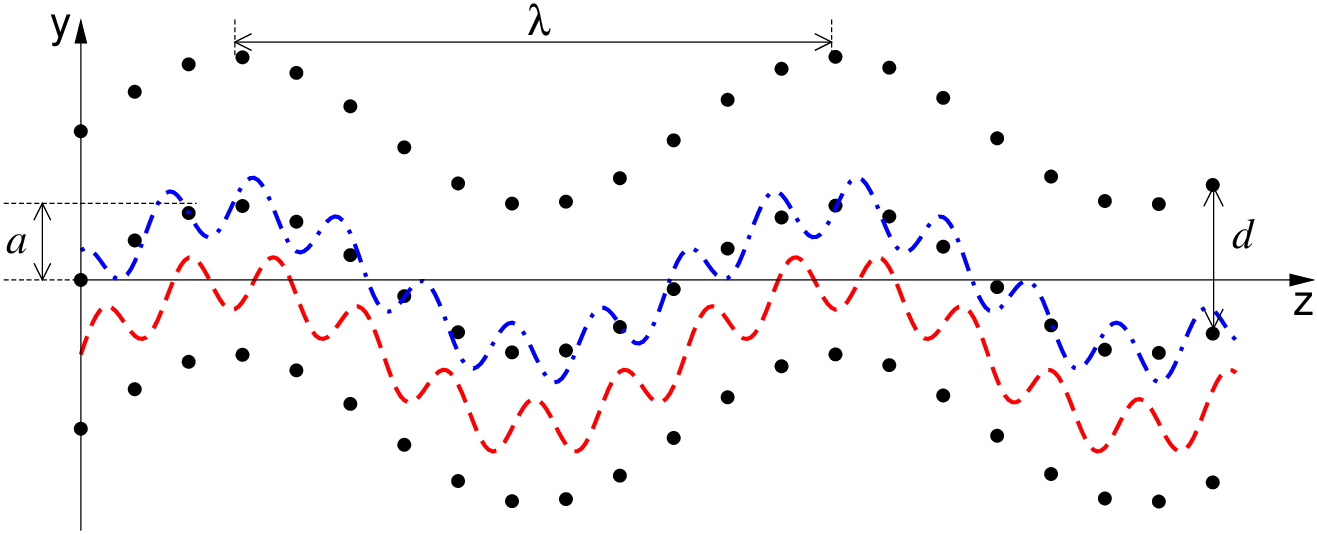}
\caption{Schematic representation of a crystalline undulator.
Circles denote the atoms belonging to 
neighboring crystallographic planes (separated by the distance $d$)
which are periodically bent.
The wavy lines represent the trajectories of channeling particles. 
A positron (dashed curve) channels in between two planes, 
whereas the electron channeling (chained curve)
occurs nearby the crystallographic plane.
The profile of periodic bending is given by
$y(z)=a\sin(2\pi z/\lambda)$, where
the period $\lambda$ and amplitude $a$ satisfy the condition
$\lambda \gg a > d$.
}
\label{figure1.fig}
\end{figure}

The mechanism of the photon emission in a crystalline undulator is 
illustrated by Fig.~\ref{figure1.fig}.  
Provided certain conditions are met, the particles, injected into the crystal,
will undergo channeling in the periodically bent channel 
\cite{KSG1998_1999}.
The trajectory of a particle contains two elements.
Firstly, there are channeling oscillations due to the action of
the interplanar potential \cite{Lindhard}.
Their typical frequency $\Om_{\ch}$ depends on the projectile energy $\E$ and 
parameters of the potential.
Secondly, there are oscillations because of the periodicity of the 
bendings, the undulator oscillations, whose
frequency is $\om_0\approx 2\pi c/\lambda$ ($c$ is the velocity of light).
The spontaneous emission 
is associated  with both of these oscillations. 
The typical frequency of the channeling radiation 
is $\om_{\ch}\approx 2\gamma^2\Om_{\ch}$ \cite{Kumakhov2}, where 
$\gamma=\E/mc^2$. 
The undulator oscillations give rise to photons with  
frequency $\om\approx 4\gamma^2\om_0/(2+p^2)$, where $p=2\pi\gamma a/\lambda$ is 
the undulator parameter.
If $\om_0 \ll \Om_{\ch}$, then the frequencies of channeling and undulator 
radiation are also well separated.
In this limit the characteristics of undulator radiation 
are practically independent on the channeling oscillations 
\cite{KSG1998_1999}, 
and the operational principle of a crystalline undulator 
is the same as for a conventional one \cite{RullhusenArtruDhez}
in which the monochromaticity of radiation is the result of 
constructive interference of the photons emitted from similar parts of 
the trajectory.

The necessary conditions, which must be met in order to treat a 
crystalline undulator as a feasible scheme for devising 
a new source of electromagnetic radiation, are as follows 
\cite{KSG1998_1999}:
\begin{eqnarray}
\begin{cases}
C =4\pi^2\E a/  \dUmax \lambda^2 < 1
&\!\!\!\! \mbox {- stable channeling},
\label{AllConditions.1}\\
d < a \ll \lambda
&\!\!\!\! \mbox {- large-amplitude regime},
\label{AllConditions.2}\\
N = L/\lambda \gg 1
&\!\!\!\! \mbox {- large number of periods},
\label{AllConditions.3}\\
L \sim \min\Bigl[\Ld(C),L_a(\om)\Bigr]
&\!\!\!\! \mbox {- account for dechanneling} \\
&         \mbox {and photon attenuation},
\label{AllConditions.4}\\
\Delta \E /\E \ll 1
&\!\!\! \!\mbox {- low radiative losses}.
\label{AllConditions.5}
\end{cases}
\end{eqnarray}
Below we present a short description of the physics lying behind these 
conditions.

A {\em stable channeling} of a projectile in a periodically bent channel 
occurs if the maximum centrifugal force $F_{\rm cf}$
is less than the maximal interplanar force $\dUmax$, i.e.
$C=F_{\rm cf}/\dUmax<1$. 
Expressing $F_{\rm cf}$ through the energy $\E$ of the projectile, the period 
and amplitude of the bending one formulates this condition 
as it is written in (\ref{AllConditions.5}).

The operation of a crystalline undulator should be considered in 
the  {\em large-amplitude regime}.
Omitting the discussion (see Ref.~\cite{KSG1998_1999,KSG2004_review}), 
we note that the limit 
$a/d>1$ accompanied by the condition $C\ll 1$  is mostly advantageous, 
since in this
case the characteristic frequencies of undulator and channeling radiation 
are well separated, so that the latter does not affect the parameters 
of the former,
whereas the intensity of undulator radiation becomes comparable or
higher than that of the channeling one \cite{KSG1998_1999,KKSG00Tot}.
A strong inequality $a\ll\lambda$, 
resulting in elastic deformation of the crystal, 
leads to moderate values of the undulator parameter $p\sim 1$ 
which ensure that the emitted radiation is of the undulator type 
rather than of the synchrotron one.

The term 'undulator' implies that the {\em number of periods, 
$N$, is large}.
Only then the emitted radiation bears the features of an undulator radiation 
(narrow, well-separated peaks in spectral-angular distribution).
This is stressed by the third condition in (\ref{AllConditions.5}).

The essential difference between  a crystalline undulator and a 
conventional one, 
based on the action of a magnetic (or electric) field 
\cite{RullhusenArtruDhez},
is that in the latter the beams of particles and photons move in vacuum 
whereas in the former \ -- \ in a crystalline medium, where
they are affected by {\em the dechanneling and the 
photon attenuation}. 
The dechanneling effect stands for a gradual increase in the 
transverse energy of a channeled particle due to inelastic collisions 
with the crystal constituents \cite{Lindhard}.
At some point the particle gains a transverse energy 
higher than the planar potential barrier and leaves the channel.
The average interval for a particle to penetrate
into a crystal until it dechannels is called the dechanneling length,
$\Ld$.
In a straight channel this quantity depends on the crystal, 
on the energy and the type of a projectile.
In a periodically bent channel there appears an additional 
dependence on the parameter $C$.
The intensity of the photon flux, which propagates through a crystal,
decreases due to the processes of absorption and scattering. 
The interval within which the intensity  decreases by a factor of $e$
is called the attenuation length,  $L_a(\om)$.
This quantity is tabulated for a number of elements 
and for a wide range of photon frequencies (see, e.g., 
Ref.~\cite{ParticleDataGroup2006}).
The forth condition in (\ref{AllConditions.5})
takes into account severe limitation of the allowed values 
of the length $L$ of a crystalline undulator due to the dechanneling 
and the attenuation.

Finally, let us comment on the last condition in (\ref{AllConditions.5}).
For sufficiently large photon energies ($\om\gtrsim 10^2$ keV)
the restriction due to the attenuation becomes less 
severe than due to the dechanneling effect \cite{KSG1998_1999,KSG2004_review}.
Then, $\Ld(C)$ introduces an upper limit on the 
length of a crystalline undulator.
Indeed, it was demonstrated \cite{SPIE1,Dechan01} that 
in the limit $L \gg \Ld$ the intensity of radiation is not 
defined by the expected number of undulator periods $L/\lambda$ but
rather is  formed in the undulator of the effective length $\sim\Ld$.
Since for an ultra-relativistic particle $\Ld \propto \E$ 
\cite{BiryukovChesnokovKotovBook,Uggerhoj_RPM2005,Baier}, 
it seems natural that to increase the effective length 
one can consider higher energies. 
However, at this point another limitation manifests itself 
\cite{KSG1998_1999,KSG00Loss}.
The coherence of an undulator radiation is only possible when
the energy loss $\Delta \E$ of the particle during its passage through the 
undulator is small, $\Delta \E\ll\E$.
This statement together with the fact, that for an ultra-relativistic 
projectile $\Delta \E$ is mainly due to the photon emission \cite{Baier}, 
leads to the 
conclusion that $L$ must be much smaller than the 
radiation length $\Lrad$, - 
the distance over which a particle converts its energy into radiation.

For a positron-based crystalline undulator 
a thorough analysis of the system  (\ref{AllConditions.5}) 
was carried out for the first time in 
Refs.~\cite{KSG1998_1999,SPIE1,KKSG00Tot,KSG00Loss,KSG2004_review}.
For a number of crystals the ranges of 
$\E$, $a$, $\lambda$ and $\om$ were established within which 
the operation of the crystalline undulator is possible.
These ranges include 
$\E=(0.5\dots 5)$ GeV, $a/d=10^1\dots10^2$,
$C=0.01\dots0.2$, 
$N\sim \Nd=\Ld/\lambda =10^1\dots10^2$, $\om\gtrsim 10^2$ keV
and are common for all the investigated crystals.
The importance of exactly this regime of operation of the  positron-based
crystalline undulator was later realized by other authors 
\cite{BellucciEtal2003,BaranovEtAl2005}.
                       
In the case of electron channeling the restrictions due to the
dechanneling effect on 
the parameters of undulator are much more severe
\cite{KSG1998_1999,KSG2004_review}. 
Therefore, it has been commonly acknowledged that the concept of an
electron-based undulator cannot be realized. 
In what follows we demonstrate, for the first time, 
that the crystalline undulator based on 
ultra-relativistic electron channeling is feasible, but it 
operates in the regime of higher beam energies than the positron-base
undulator.

\begin{figure}[ht]
\includegraphics*[width=8cm,height=5cm]{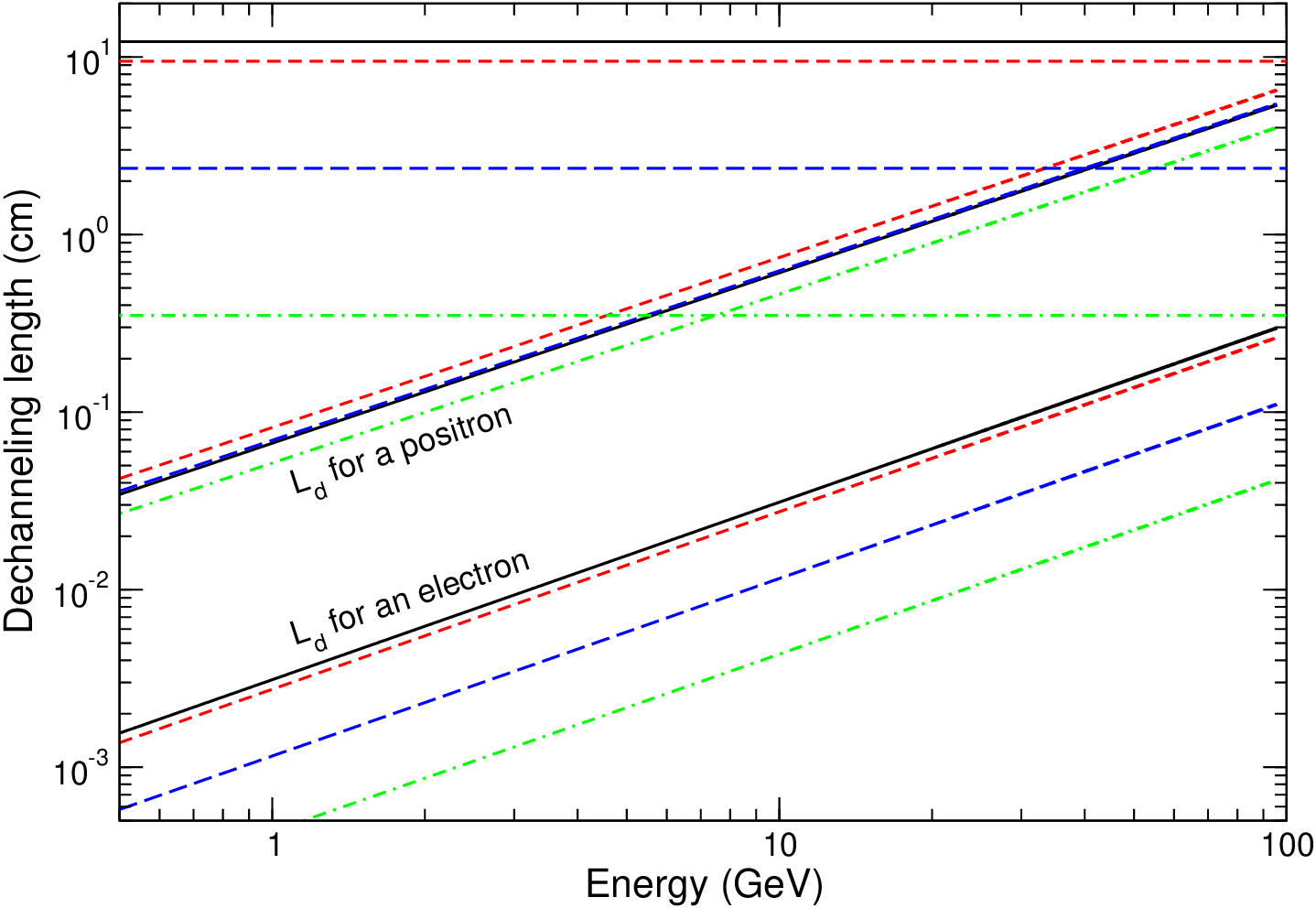}
\caption{
Positron and electron 
dechanneling lengths in the straight channels 
versus $\E$.
Solid, dashed, long-dashed and chain lines correspond to
C (111), Si (111), Ge (111) and W (110). 
The horizontal lines show the radiation lengths
\cite{comment}.  
}
\label{figure2.fig}
\end{figure}

It is important to note that for negative and for positive projectiles
the dechanneling occurs in different regimes.
Positrons, being repulsed by the interplanar potential,
channel in the regions between two neighboring planes,
whereas electrons channel in close vicinity of ion planes 
(see Fig.~\ref{figure1.fig}).
Therefore, the number of collisions with the crystal constituents 
is much larger for electrons  and they dechannel faster.
Fig.~\ref{figure2.fig}, which presents the dependences of $\Ld$ on $\E$ 
for planar channeling of positrons and electrons in various straight crystals,
illustrates this statement \cite{comment}.
It is seen that for all energies the dechanneling length for
$e^{+}$ exceeds that for $e^{-}$ by more than an order of magnitude.
Such a large difference 
(consistent with the experimental
\cite{Uggerhoj1980} and other theoretical \cite{Kumakhov2} data)
is the reason why the crystalline undulator problem has been analyzed, 
so far, only for positrons.

\begin{figure}[ht]
\includegraphics*[scale=0.37]{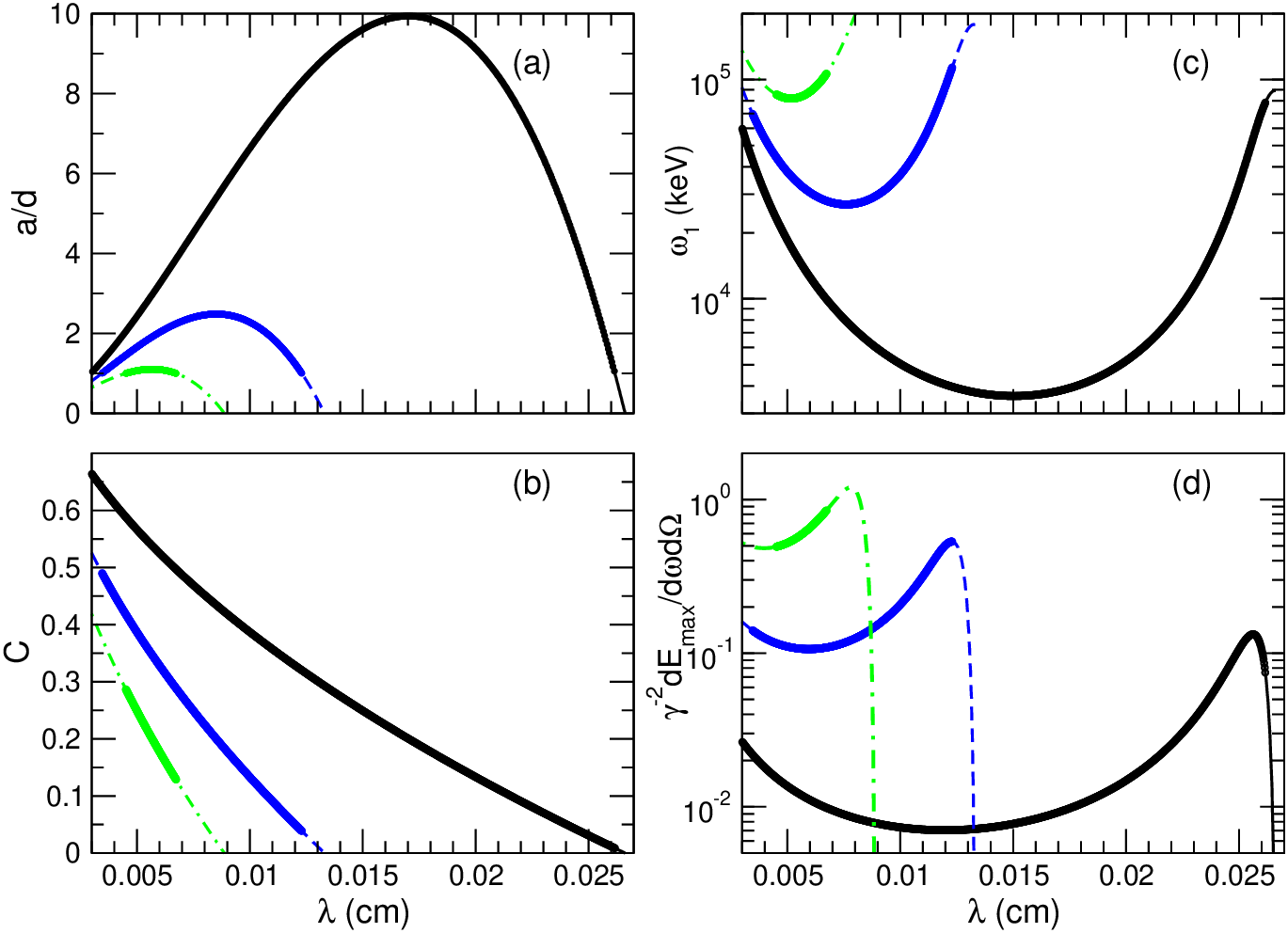}
\caption{
Parameters of the undulator and undulator radiation as functions of 
$\lambda$ for $50$ GeV electron channeling 
in Si along the (111) crystallographic planes.
In each graph solid, dashed and chain curves correspond to the fixed 
numbers of undulator periods within the dechanneling length:
$\Nd =5, 10, 15$, respectively.
Thick parts of the curves mark the regions where $a/d\geq 1$.
}
\label{figure3.fig}
\end{figure}

As mentioned, 
a {\em positron}-based undulator is feasible for $\E\lesssim 5$ GeV.
For these energies, see Fig.~\ref{figure2.fig}, 
 the radiation length greatly exceeds $\Ld$ 
(or, in other words, $\Delta\E \ll \E$), 
and it is possible to achieve $N\sim10\dots10^2$ within $\Ld$
\cite{KSG2004_review,SPIE1}.
The corresponding values of the undulator period are 
$\lambda=10^{-4}\dots10^{-2}$ cm, i.e. exactly the interval 
to which the {\em electron} dechanneling lengths belong. 
Therefore, for $0.5\dots 5$ GeV electrons the number of undulator periods 
is $\sim 1$, thus indicating 
that this system is not an undulator.

\begin{figure*}[ht]
\includegraphics*[scale=0.75]{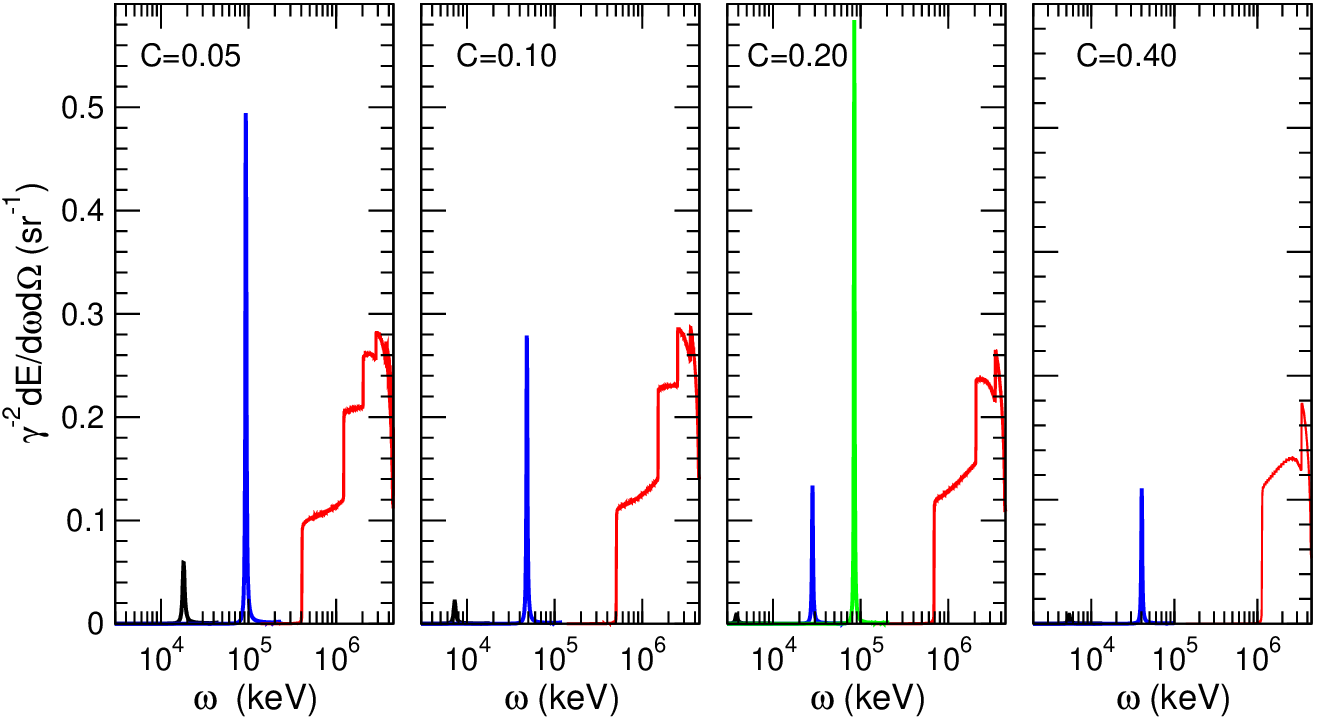}
\caption{
Spectra of the undulator (narrow peaks) and channeling (wide peaks) radiation 
emitted in the forward direction for $50$ GeV 
electron channeling in Si (111) and 
for different values of the parameter $C$.
The undulator radiation: 
the first (i.e., the leftest) narrow peaks correspond to $\Nd =5$
(in the figures for C=0.20 and C=0.40 they are hardly visible), 
the second peaks - to $\Nd =10$, and the
third peak (only in the figure for C=0.20) - to $\Nd =15$.
The crystal length is chosen as $L\approx4\Ld(C)$.
}
\label{figure4.fig}
\end{figure*}

However,  Fig.~\ref{figure2.fig} suggests that the electron-based
undulator can be discussed for higher energies, 
$\E=10\dots10^2$ GeV, where $\Ld$ is large enough to 
ensure $\Nd\gg 1$ but still is much lower than $\Lrad$.
To demonstrate the feasibility of such an undulator
one must carry out the analysis of other conditions from 
(\ref{AllConditions.5}) and establish the ranges 
of $a$, $\lambda$ and $\om$ within which the operation of the
electron-based crystalline undulator is possible.
In Figs.~\ref{figure3.fig} and \ref{figure4.fig} 
we present the results of such an analysis performed for $50$ GeV electron 
channeling in Si (111).

Figs.~\ref{figure3.fig}(a,b)
present the ranges of parameters of the electron-based undulator.
In Fig.~\ref{figure3.fig}(a) the ratio $a/d$ versus $\lambda$ 
is shown for fixed values  of undulator periods within the dechanneling 
length, i.e. for $\Nd=\Ld(C)/\lambda=const$
(the curves correspond to $\Nd=5, 10, 15$ and this 
is also valid for other graphs in the figure).
This graph illustrates the ranges of $a$, $\lambda$ and $\Nd$
within which the second and third conditions from (\ref{AllConditions.5}) 
are met.
Fig.~\ref{figure3.fig}(b)
presents the dependences $C(\lambda)$
and illustrates the fulfillment of the condition for the stable channeling.
Figs.~\ref{figure3.fig}(a,b) suggest that the undulator can be devised 
for $a=2\dots 20$ \AA, $\lambda=10\dots10^2\, \mu$m, 
which are close to parameters of a positron-based undulator 
\cite{KSG2004_review,SPIE1,PECU}.
Therefore, to construct an electron-based undulator one can consider 
the methods proposed earlier:
propagation of an acoustic wave \cite{Kaplin,KSG1998_1999}, or
the use of a graded composition of different layers
\cite{MikkelsenUggerhoj2000,PECU}, or
periodic mechanical deformation of the crystalline structure 
\cite{BellucciEtal2003,PECU}.

Figs.~\ref{figure3.fig}(c,d) 
present the parameters of the undulator radiation, 
- the energy of fundamental harmonic, $\om_1=8\pi\gamma^2 c \lambda^{-1}/(2+p^2)$, 
and the peak value of the spectral distribution 
$d^3 E_{\max}/\hbar d\om\, d\Om$ (scaled by the factor $\gamma^{-2}$)  
of the energy emitted in the forward direction at $\om = \om_1$ 
- as functions of $\lambda$.  
To calculate $d^3 E_{\max}/\hbar d\om\, d\Om$ we used the formalism, 
developed in \cite{SPIE1} to describe the undulator radiation
in presence of the dechanneling and the photon attenuation 
(i.e. the decrease in the intensity of the photon flux in a crystal).
For each $\lambda$ the crystal length was chosen as
$L\approx4\Ld(C)$.
This value is close to the optimal length of the undulator, which
ensures the highest yield of the photons for given $C$, $\E$ and $\om_1$
\cite{SPIE1}. 

Figs.~\ref{figure3.fig}(c,d) 
show that both the photon energy and the intensity of undulator radiation can
be varied over wide ranges (note the log scale).
However, it is important to compare these quantities with 
the characteristics of the channeling radiation.
This is done in  Fig.~\ref{figure4.fig}, 
where the spectral distributions of undulator and channeling 
radiation in the forward direction are presented for the indicated 
values of $C$
(using which in Fig.~\ref{figure3.fig}(a,b) 
one finds the values of $a/d$ and $\lambda$).
The spectra were calculated  with the account 
for the dechanneling and the photon attenuation \cite{SPIE1}.
To calculate the spectrum of channeling radiation we, first,
calculated the spectra for individual trajectories 
(using the P\"oschl-Teller model \cite{Baier} 
for the interplanar potential 
acting on an electron), 
corresponding to stable channeling for given $C$.
Then, 
the averaging procedure was applied to calculate the spectra
(see Refs.~\cite{KKSG00Tot} for the details).

The graphs in Fig.~\ref{figure4.fig} 
clearly demonstrate that by tuning the parameters of bending
it is possible to separate the frequencies of the
undulator radiation from those of the channeling radiation, and
to make the intensity of the former comparable 
or higher than of the latter.

In summary, we have demonstrated that it is 
feasible to devise an undulator based on the channeling effect of
a bunch of ultra-relativistic electrons through a periodically bent crystal.
An electron-based undulator operates in the regime of higher 
energies of projectiles
than a positron-based one.
The present technologies are sufficient to
achieve the necessary conditions to construct the 
undulator and to create, on its basis,
powerful radiation sources in the $\gamma$-region of the spectrum.
As in the positron case \cite{KSG1998_1999} it is meaningful
to explore the idea of a $\gamma$-laser 
by means of an electron-based 
undulator.

This work has been supported by the European Comission 
(the PECU project, Contract No. 4916 (NEST)).


\end{document}